\documentclass[final,3p,twocolumn,times]{elsarticle}

\usepackage{graphicx}
\usepackage{amsmath}
\usepackage{amssymb}
\usepackage{amsthm}
\usepackage{bm}
\usepackage{color}
\usepackage{stmaryrd}
\usepackage{mathrsfs}
\usepackage{array}
\usepackage{algorithm}
\usepackage{algpseudocode}
\usepackage{multirow}
\usepackage{url}
\usepackage{lineno}

\SetSymbolFont{stmry}{bold}{U}{stmry}{m}{n}

\numberwithin{equation}{section}
\newcolumntype{P}[1]{>{\centering\arraybackslash}p{#1}}

\journal{}

\begin{document}

\begin{frontmatter}

\title{Fluid-structure interaction modeling of blood flow in the pulmonary arteries using the unified continuum and variational multiscale formulation} 

\author{Ju Liu}
\ead{liuju@stanford.edu,liujuy@gmail.com}

\author{Weiguang Yang}
\ead{wgyang@stanford.edu}

\author{Ingrid S. Lan}
\ead{ingridl@stanford.edu}

\author{Alison L. Marsden}
\ead{amarsden@stanford.edu}

\address{Department of Pediatrics (Cardiology), Department of Bioengineering, and Institute for Computational and Mathematical Engineering, Stanford University, Clark Center E1.3, 318 Campus Drive, Stanford, CA 94305, USA}

\begin{abstract}
In this work, we present a computational fluid-structure interaction  (FSI) study for a healthy patient-specific pulmonary arterial tree using the unified continuum and variational multiscale (VMS) formulation we previously developed. The unified framework is particularly well-suited for FSI, as the fluid and solid sub-problems are addressed in essentially the same manner and can thus be uniformly integrated in time with the generalized-$\alpha$ method. In addition, the VMS formulation provides a mechanism for large-eddy simulation in the fluid sub-problem and pressure stabilization in the solid sub-problem. The FSI problem is solved in a quasi-direct approach, in which the pressure and velocity in the unified continuum body are first solved, and the solid displacement is then obtained via a segregated algorithm and prescribed as a boundary condition for the mesh motion. Results of the pulmonary arterial FSI simulation are presented and compared against those of a rigid wall simulation.
\end{abstract}


\end{frontmatter}

\section{Introduction}
\label{sec:introduction}
We recently derived a unified continuum formulation based on the Gibbs free energy in order to construct a well-behaved continuum model in both compressible and incompressible regimes \cite{Liu2018}. This modeling approach naturally recovers important continuum models, including viscous fluids and hyperelastic solids. Importantly, it bridges previously diverging approaches in computational fluid dynamics (CFD) and computational solid dynamics (CSD). The residual-based VMS formulation can be applied to the unified continuum body. It yields a large-eddy simulation procedure for the incompressible Navier-Stokes equations \cite{Bazilevs2007a}, which performs equally well for laminar, transitional, and fully turbulent flows \cite{Hughes2001,Liu2020}. On the other hand, when applied to the hyperelastic model, it leads to a numerical formulation for finite elasticity that allows equal-order interpolation of all fields. This is particularly beneficial for problems with complex geometries and bears similarity to some recent works \cite{Scovazzi2016,Rossi2016,Gil2014,Masud2013}. In our opinion, the unified concept gives rise to promising opportunities for designing new numerical methodologies. Recent advances include the development of a provably energy-stable scheme for incompressible finite elasticity \cite{Liu2019a} and preconditioning techniques for both solids \cite{Liu2019} and fluids \cite{Liu2020}. The benefit of the unified modeling framework is further evident in the realm of multiphysics coupled problems. Since the CFD and CSD implementations only differ in constitutive routines, monolithic FSI coupling is dramatically simplified. Furthermore, in comparison with conventional FSI modeling approaches \cite{Bazilevs2013,Yan2016,Bazilevs2010,Takizawa2012}, the new framework allows one to simulate structural dynamics with a Poisson's ratio up to $0.5$, using either the multiscale/stabilized formulation or inf-sup stable methods. Since soft tissues typically exhibit nearly incompressible behavior under physiologic loading \cite{Humphrey2013}, the proposed FSI modeling framework is extremely favorable for computational biomechanics and cardiovascular hemodynamics.

In this work, we present a suite of FSI modeling techniques for cardiovascular applications. In addition to the unified FSI modeling framework, we discuss mesh generation from medical image data as well as a modular approach for implicit coupling of lumped parameter network (LPN) models with the three-dimensional (3D) domain \cite{Moghadam2013}. The efficacy of the proposed methodology is demonstrated through a numerical study in the pulmonary arteries of a pediatric patient. The FSI results are directly compared to those of a rigid wall simulation.

\section{The unified continuum formulation for fluid-structure interaction}
\label{sec:unfiied-continuum-formulation}
In this section, we present the governing equations for the FSI problem using the arbitrary Lagrangian-Eulerian (ALE) method \cite{Bazilevs2013,Scovazzi2007}. Here, and in what follows, we use superscripts $f$, $s$, and $m$ to indicate quantities related to the fluid, solid, and ALE mesh motion in the fluid sub-domain.

\subsection{Kinematics on moving domains}
We first consider the domain occupied by the continuum body in the referential frame $\Omega_{\bm \chi} \subset \mathbb R^3$, an open and bounded set. For FSI problems, $\Omega_{\bm \chi}$ admits a non-overlapping subdivision, $\overline{\Omega}_{\bm \chi} = \overline{ {\Omega}_{\bm \chi}^{f} \cup {\Omega}_{\bm \chi}^{s} }$, $\emptyset = \Omega_{\bm \chi}^{f} \cap \Omega_{\bm \chi}^{s}$, in which $\Omega^{f}_{\bm \chi}$ and $\Omega^{s}_{\bm \chi}$ represent the sub-domains occupied by the fluid and solid, respectively. Following the notation used in \cite{Liu2018}, the referential-to-Eulerian map at time $t$ is denoted $\hat{\bm \varphi}_t(\cdot) = \hat{\bm \varphi}(\cdot, t)$ and maps $\Omega_{\bm \chi}$ to $\Omega_{\bm x}(t) = \hat{\bm \varphi}\left(\Omega_{\bm \chi}, t\right)$. We wish to think of $\Omega_{\bm x}(t)$ as the current `spatial' domain where the fluid mechanics problem can be conveniently formulated. Correspondingly, the current configuration admits a subdivision, $\overline{\Omega}_{\bm x}(t) = \overline{\Omega^{f}_{\bm x}(t) \cup \Omega^{s}_{\bm x}(t)}$, $\emptyset = \Omega^{f}_{\bm x}(t) \cap \Omega^{s}_{\bm x}(t)$. Conceptually, $\Omega_{\bm \chi}$ is fixed in time and is associated with a computational mesh. Therefore, $\hat{\bm \varphi}$ describes the motion of the mesh, and we can correspondingly define the mesh displacement and velocity as
\begin{align}
& \hat{\bm U}^m := \hat{\bm \varphi}(\bm \chi, t) - \hat{\bm \varphi}(\bm \chi, 0) = \hat{\bm \varphi}(\bm \chi, t) - \bm \chi, \\
\label{eq:mesh-disp-velo}
& \hat{\bm V}^m := \left. \frac{\partial \hat{\bm \varphi}}{\partial t} \right|_{\bm \chi} = \left. \frac{\partial \hat{\bm U}^m}{\partial t} \right|_{\bm \chi}.
\end{align}
One may conveniently push them forward to the current configuration as $\hat{\bm u}^m := \hat{\bm U}^m \circ \hat{\bm \varphi}_t^{-1}$ and $\hat{\bm v}^m := \hat{\bm V}^m \circ \hat{\bm \varphi}_t^{-1}$. 

The initial position of point $\bm x \in \Omega_{\bm x}(t)$ is denoted as $\bm X \in \Omega_{\bm X}(t)$, where $\Omega_{\bm X}(t)$ is the Lagrangian domain. The smooth Lagrangian-to-Eulerian map at time $t$ is denoted $\bm \varphi_t(\cdot) = \bm \varphi(\cdot, t)$ and maps $\Omega_{\bm X}(t)$ to $\Omega_{\bm x}(t)$. Then the displacement, velocity, deformation gradient, the Jacobian determinant, the right Cauchy-Green tensor of the material particle initially located at $\bm X$ are defined as
\begin{align*}
& \bm U := \bm \varphi(\bm X, t) - \bm \varphi(\bm X,0) = \bm \varphi(\bm X, t) - \bm X, \\
& \bm V := \left. \frac{\partial \bm \varphi}{\partial t}\right|_{\bm X}= \left. \frac{\partial \bm U}{\partial t}\right|_{\bm X} = \frac{d\bm U}{dt}, \\
& \bm F:= \frac{\partial \bm \varphi}{\partial \bm X}, \quad
J := \textup{det}\left(\bm F \right), \quad \bm C := \bm F^T \bm F.
\end{align*}
The displacement and velocity can be similarly pushed forward to the current configuration as $\bm u := \bm U \circ \bm \varphi_t^{-1}$ and $\bm v := \bm V \circ \bm \varphi_t^{-1}$. We also introduce the distortional parts of $\bm F$ and $\bm C$ as
\begin{align*}
\tilde{\bm F} := J^{-\frac13} \bm F, \quad \tilde{\bm C} := J^{-\frac23} \bm C.
\end{align*}

\subsection{Balance and mesh motion equations}
\label{subsec:balance-equations}
We invoke Stokes' hypothesis and further consider the isothermal condition on the continuum body, allowing the energy equation to be decoupled from the mechanical system. The FSI system can thus be viewed as a two-component continuum body governed by the following momentum and mass balance equations,
\begin{align*}
\bm 0 &= \rho(p) \left.  \frac{\partial \bm v}{\partial t}\right|_{\bm \chi}  + \rho(p) \left( \bm v - \hat{\bm v}^m \right) \cdot \nabla_{\bm x} \bm v - \nabla_{\bm x} \cdot \bm \sigma_{dev} + \nabla_{\bm x} p \nonumber \\
& \hspace{3mm} - \rho(p) \bm b, \\
0 &= \left. \beta_{\theta}(p)\frac{\partial p}{\partial t}\right|_{\bm \chi} + \beta_{\theta}(p) \left( \bm v - \hat{\bm v}^m \right) \cdot \nabla_{\bm x} p + \nabla_{\bm x} \cdot \bm v,
\end{align*}
which are posed in $\Omega_{\bm x}(t)$. In the above equations, $\rho$ is the density, $p$ is the pressure, $\bm \sigma_{dev}$ is the deviatoric part of the Cauchy stress, $\bm b$ is the body force per unit mass, and $\beta_{\theta}$ is the isothermal compressibility factor. The constitutive laws of the material are dictated by the Gibbs free energy $G(\tilde{\bm C}, p)$, which was previously shown to adopt a decoupled structure \cite[p.~559]{Liu2018},
\begin{align*}
G(\tilde{\bm C}, p) = G_{ich}(\tilde{\bm C} ) + G_{vol}(p),
\end{align*}  
where $G_{ich}$ and $G_{vol}$ represent the isochoric and volumetric parts of the free energy, respectively. Given the free energy, the constitutive relations can be written as
\begin{align*}
& \rho(p) := \left( \frac{d G_{vol}}{d p} \right)^{-1}, \hspace{1mm} \beta_{\theta}(p) := \frac{1}{\rho} \frac{d\rho}{d p} = -\frac{d^2 G_{vol}}{d p^2} / \frac{d G_{vol}}{d p}, \displaybreak[2] \\
& \bm \sigma_{dev} := J^{-1} \tilde{\bm F} \left( \mathbb P : \tilde{\bm S} \right) \tilde{\bm F}^T + 2\bar{\mu} \textup{dev}[\bm d], \displaybreak[2] \\
& \mathbb P := \mathbb I - \frac13 \bm C^{-1} \otimes \bm C, \quad \tilde{\bm S} := 2 \frac{\partial \left(\rho_0 G \right) }{\partial \tilde{\bm C} }, \displaybreak[2] \\
& \bm d := \frac12 \left(\nabla_{\bm x} \bm v + \nabla_{\bm x} \bm v^T \right),
\end{align*}
where $\mathbb I$ is the fourth-order identity tensor, and $\rho_0$ is the density in the Lagrangian domain. 

In the solid sub-domain, we consider a purely elastic material and choose the referential configuration to be identical to the Lagrangian configuration. Consequently, the balance equations in $\Omega^s_{\bm x}(t)$ can be stated as
\begin{align*}
\bm 0 &= \rho^s(p^s) \left.  \frac{\partial \bm v^s}{\partial t}\right|_{\bm \chi = \bm X} - \nabla_{\bm x} \cdot \bm \sigma^s_{dev} + \nabla_{\bm x} p^s - \rho^s(p^s) \bm b, \\
0 &= \left. \beta^s_{\theta}(p^s)\frac{\partial p^s}{\partial t}\right|_{\bm \chi = \bm X} + \nabla_{\bm x} \cdot \bm v^s.
\end{align*}
In the fluid sub-domain, the free energy contains no mechanical contribution, so $\bm \sigma_{dev}^f = 2\bar{\mu}\textup{dev}[\bm d]$. We further assume incompressible flow, which implies $\rho^f(p^f) = \rho^f$ and $\beta_{\theta}^f = 0$. The balance equations in $\Omega^f_{\bm x}(t)$ are then
\begin{align*}
\bm 0 &= \rho^f \left.  \frac{\partial \bm v^f}{\partial t}\right|_{\bm \chi}  + \rho^f \left( \bm v^f - \hat{\bm v}^m \right) \cdot \nabla_{\bm x} \bm v^f - \nabla_{\bm x} \cdot \bm \sigma^f_{dev} + \nabla_{\bm x} p^f \nonumber \\
& \hspace{3mm} - \rho^f \bm b,\\
0 &= \nabla_{\bm x} \cdot \bm v^f.
\end{align*}
In this work, we use the pseudo-linear-elasticity algorithm to model the ALE mesh motion \cite{Bazilevs2008,Johnson1994}. Consider a time instant $\tilde{t} < t$, which is often chosen to be the previous time step in numerical computations. Given the identity $\hat{\bm \varphi}(\bm \chi, t) = \hat{\bm \varphi}(\bm \chi, \tilde{t}) + \hat{\bm U}(\bm \chi, t ) - \hat{\bm U}(\bm \chi, \tilde{t})$, we introduce $\tilde{\bm u}^m \left( \hat{\bm \varphi}(\bm \chi, \tilde{t}) ,t \right) := \hat{\bm U}(\bm \chi, t) - \hat{\bm U}(\bm \chi, \tilde{t})$. The mesh velocity $\hat{\bm v}^m$ is then completely determined by $\tilde{\bm u}^m(\tilde{\bm x}, t)$ and the relation in \eqref{eq:mesh-disp-velo}. The mesh motion is solved via the following linear elastostatic problem posed in $\Omega^{f}_{\bm x}(\tilde{t})$,
\begin{align*}
& \nabla_{\tilde{\bm x}}  \cdot \left( \mu^m \left( \nabla_{\tilde{\bm x}} \tilde{\bm u}^m + \left(\nabla_{\tilde{\bm x}} \tilde{\bm u}^m\right)^T  \right) + \lambda^m  \nabla_{\tilde{\bm x}} \cdot \tilde{\bm u}^m \bm I \right) = \bm 0. 
\end{align*}
The boundary of the fluid sub-domain can be decomposed into the luminal, inlet, and outlet surfaces. On the luminal surface, the mesh motion follows the motion of the solid body and is therefore subject to a Dirichlet boundary condition; on the inlet and outlet surfaces, we prescribe homogeneous Dirichlet boundary conditions to fix the mesh. Furthermore, to enhance the robustness of the mesh moving algorithm, the Lam\'e parameters $\mu^m$ and $\lambda^m$ are chosen to be proportional to the inverse of the Jacobian determinant of the element mapping \cite{Johnson1994,Bazilevs2013}.

\section{Numerical formulation}

\subsection{Solid sub-problem}
\label{subsec:solid-spatial-formulation}
Let $\mathcal S^s_{\bm u}$, $\mathcal S^{s}_{\bm v}$, and $\mathcal S^{s}_{p}$ denote the finite dimensional trial solution spaces for the solid displacement, velocity, and pressure in the current solid sub-domain, respectively; let $\mathcal V^s_{\bm v}$, and $\mathcal V^s_{p}$ represent the test function spaces; let $\Gamma^s_{\bm x,h}(t)$ denote the Neumann part of the solid boundary with traction $\bm h^s$ prescribed. The spatial discretization for the solid body is based on the variational multiscale formulation \cite{Liu2018}, which is stated as follows: Find $\left\lbrace  \bm u_h^s(t), p_h^s(t), \bm v_h^s(t)\right\rbrace \in \mathcal S_{\bm u}^s \times \mathcal S_{p}^s \times \mathcal S_{\bm v}^s$ such that for $\forall  \left\lbrace \bm w^s, w^s\right\rbrace \in \mathcal V_{\bm v}^s \times \mathcal V_{p}^s$,
\begin{align*}
& \bm 0 = \frac{d\bm u_h^s}{dt} - \bm v_h^s, \displaybreak[2] \\
& 0 = \int_{\Omega_{\bm x}^s(t)} \bm w^s \cdot \rho^s(p^s_h) \frac{d\bm v_h^s}{dt} d\Omega_{\bm x} + \int_{\Omega_{\bm x}^s(t)} \nabla_{\bm x} \bm w^s : \bm \sigma^s_{dev}(\bm u^s_h) d\Omega_{\bm x} \nonumber \\  
&  \hspace{3mm} - \int_{\Omega_{\bm x}^s(t)} \nabla_{\bm x} \cdot \bm w^s p_h^s d\Omega_{\bm x} - \int_{\Omega_{\bm x}^s(t)}\bm w^s \cdot \rho^s(p^s_h) \bm b d\Omega_{\bm x} \nonumber \\
& \hspace{3mm} -\int_{\Gamma_{\bm x,h}^{s}(t)}\bm w^s \cdot \bm h^s d\Gamma_{\bm x},\\
& 0 = \int_{\Omega_{\bm x}^s(t)} w^s \left( \beta_{\theta}^s(p^s_h) \frac{dp_h^s}{dt} + \nabla_{\bm x} \cdot \bm v_h^s \right) d\Omega_{\bm x} \nonumber \\
& \hspace{3mm} - \int_{\Omega_{\bm x}^{\prime s}(t)} \nabla_{\bm x} w^s \cdot \bm v^{s\prime} d\Omega_{\bm x}, \\
& \bm v^{s\prime} := -\bm \tau_M^s \left( \rho^s(p_h^s) \frac{d\bm v_h^s}{dt} - \nabla_{\bm x} \cdot \bm \sigma^s_{dev}(\bm u^s_h) + \nabla_{\bm x} p^s_h - \rho^s(p^s_h) \bm b \right).
\end{align*}
In the above formulation, the parameter $\bm \tau_M^s$ is associated with the subgrid-scale models and is defined as 
\begin{align*}
\bm \tau_M^s = \tau_M^s \bm I, \quad \tau_M^s = c_m \frac{\Delta x}{c\rho^s},
\end{align*}
where $\Delta x$ is the diameter of the circumscribing sphere of the tetrahedral element, $c$ is the maximum wave speed in the solid material, and $c_m$ is a non-dimensional scalar \cite{Scovazzi2016}.

\subsection{Mesh motion of the fluid sub-domain}
Let $\mathcal S^m_{\tilde{\bm u}}$ denote the trial solution space of the mesh displacement $\tilde{\bm u}^m_h$ defined on the domain $\Omega_{\bm x}^{f}(\tilde{t})$, and let $\mathcal V^m_{\tilde{\bm u}}$ denote the corresponding test function space. The variational formulation of the problem is stated as follows. Find $\tilde{\bm u}^m_h \in \mathcal S^m_{\tilde{\bm u}}$ such that for $\forall \tilde{\bm w}^m \in \mathcal V^m_{\tilde{\bm u}}$,
\begin{align*}
\int_{\Omega_{\bm x}^f(\tilde{t})} \nabla^s_{\tilde{\bm x}} \tilde{\bm w}^m : \left( 2\mu^m \nabla^s_{\tilde{\bm x}} \tilde{\bm u}^m_h \right) + \nabla_{\tilde{\bm x}} \cdot \tilde{\bm w}^m \lambda^m \nabla_{\tilde{\bm x}} \cdot \tilde{\bm u}^m_h d\Omega_{\bm x} = 0.
\end{align*}

\subsection{Fluid sub-problem}
\label{subsec:fluid-spatial-formulation}
Let $\mathcal S^f_{\bm v}$ and $\mathcal S^f_{p}$ denote the trial solution space of the fluid velocity and pressure; let $\mathcal V^f_{p}$ and $\mathcal V^f_{\bm v}$ be the test function spaces; let $\Gamma^f_{\bm x,h}(t)$ denote the Neumann part of the fluid boundary with traction $\bm h^f$ prescribed. The VMS formulation for the fluid sub-problem can be stated as follows. Find $\left\lbrace p_h^f(t), \bm v_h^f(t) \right\rbrace \in \mathcal S^f_{p} \times \mathcal S^f_{\bm v}$ such that for $\forall  \left\lbrace \bm w^f, w^f\right\rbrace \in \mathcal V_{\bm v}^f \times \mathcal V_{p}^f$,
\begin{align*}
& 0 = \int_{\Omega_{\bm x}^f(t)} \bm w^f \cdot \left( \left. \rho^f \frac{\partial \bm v_h^f}{\partial t} \right|_{\bm \chi} + \rho^f \left(\bm v_h^f - \hat{\bm v}^m_h \right) \cdot \nabla_{\bm x} \bm v_h^f \right) d\Omega_{\bm x} \nonumber \displaybreak[2] \\
& - \int_{\Omega_{\bm x}^f(t)} \nabla_{\bm x} \cdot \bm w^f p_h^f d\Omega_{\bm x} + \int_{\Omega_{\bm x}^f(t)} \nabla_{\bm x} \bm w^f : \bm \sigma_{dev}^f(\bm v^f_h) d\Omega_{\bm x} \nonumber \displaybreak[2] \\
& - \int_{\Omega_{\bm x}^f(t)} \bm w^f \cdot \rho^f \bm b d\Omega_{\bm x} - \int_{\Gamma^{f}_{\bm x,h}(t)} \bm w^f \cdot \bm h^f d\Gamma_{\bm x} \nonumber \displaybreak[2] \\
& - \int_{\Omega_{\bm x}^{\prime f}(t)} \nabla_{\bm x} \bm w^f : \left( \rho^f \bm v^{f\prime} \otimes \left( \bm v_h^f - \hat{\bm v}^m_h \right) \right) d\Omega_{\bm x} \nonumber \displaybreak[2] \\
& + \int_{\Omega_{\bm x}^{\prime f}(t)} \nabla_{\bm x} \bm v^f_h : \left( \rho^f \bm w^f \otimes \bm v^{f\prime} \right) - \nabla_{\bm x} \bm w^f : \left( \rho^f \bm v^{f\prime} \otimes \bm v^{f\prime} \right) d\Omega_{\bm x} \nonumber \displaybreak[2] \\
& - \int_{\Omega_{\bm x}^{\prime f}(t)} \nabla_{\bm x} \cdot \bm w^f p^{f\prime} d\Omega_{\bm x}, \displaybreak[2] \\
& 0 = \int_{\Omega_{\bm x}^f(t)} w^f \nabla_{\bm x} \cdot \bm v_h^f d\Omega_{\bm x}  - \int_{\Omega_{\bm x}^{\prime f}(t)} \nabla_{\bm x} w^f \cdot  \bm v^{f\prime} d\Omega_{\bm x}, \displaybreak[2] \\
& \bm v^{f\prime} := -\bm \tau_{M}^f \Big( \left. \rho^f \frac{\partial \bm v_h^f}{\partial t} \right|_{\bm \chi} + \rho^f \left( \nabla_{\bm x} \bm v_h^f \right) \left(\bm v_h^f - \hat{\bm v}^m_h \right) + \nabla_{\bm x} p_h^f \nonumber \\
& \hspace{8mm} - \nabla_{\bm x} \cdot \bm \sigma^f_{dev}(\bm v^f_h) - \rho^f \bm b \Big), \displaybreak[2] \\
& p^{f\prime} := -\tau^f_C \nabla_{\bm x} \cdot \bm v_h^f, \displaybreak[2] \\
& \bm \tau^f_M := \tau^f_M \bm I, \displaybreak[2] \\
& \tau^f_M := \frac{1}{\rho^f}\left( \frac{C_T}{\Delta t^2} + \left(\bm v_h^f - \hat{\bm v}^m_h \right) \cdot \bm G \left(\bm v_h^f - \hat{\bm v}^m_h \right)  \right. \displaybreak[2] \\
& \hspace{8mm} \left. + C_I \left( \frac{\bar{\mu}}{\rho^f} \right)^2 \bm G : \bm G \right)^{-\frac12}, \displaybreak[2] \\
& \tau^f_C := \frac{1}{\tau_M \textup{tr}\bm G}, \displaybreak[2] \\
& G_{ij} := \sum_{k=1}^{3} \frac{\partial \xi_k}{\partial x_i} M_{kl} \frac{\partial \xi_l}{\partial x_j}, \displaybreak[2] \\
& \bm M = [ M_{kl} ] = \frac{\sqrt[3]{2}}{2}\begin{bmatrix}
2 & 1 & 1 \\
1 & 2 & 1 \\
1 & 1 & 2
\end{bmatrix}, \displaybreak[2] \\
& \bm G : \bm G := \sum_{i,j=1}^{3} G_{ij} G_{ij}, \displaybreak[2] \\
& \textup{tr}\bm G := \sum_{i=1}^{3} G_{ii}.
\end{align*}
In the above, $\bm \xi= \left\lbrace \xi_i \right\rbrace_{i=1}^{3}$ represents the natural coordinates in the parent domain. The values of $C_I$ and $C_T$ are chosen to be $36$ and $4$ in this study. $\bm M$ is introduced for simplex elements to give a node-numbering-invariant definition of $\tau^f_M$ and $\tau^f_C$ \cite{Danwitz2019}.

\subsection{Boundary conditions}
For the solid sub-problem, we prescribe homogeneous Dirichlet boundary conditions on the annulus surfaces at the inlet and outlets and zero traction on the external surface of the arterial wall. 

For the fluid sub-problem, we prescribe the no-slip boundary condition on the luminal surface. On the inlet surface, we prescribe a Poiseuille velocity profile scaled by a periodic volumetric flow waveform. A special mapping technique introduced in \cite{Takizawa2010} is utilized to generate the inflow profile. To achieve physiological flows and pressures, we couple LPN models to the outlet surfaces as traction boundary conditions mimicking the effect of the downstream circulation. For each outlet surface $\Gamma^k_{\mathrm{out}}$ with unit outward normal vector $\bm n^k$, where $k$ is the outlet surface index, we prescribe
\begin{align}
\label{eq:outflow_bc}
\bm h^f = -P^k(t) \bm n^k + \beta \rho^f \left\lbrace \left( \bm v^f_h - \hat{\bm v}^m_h \right) \cdot \bm n^k \right\rbrace_{-} \bm v^f_h,
\end{align}
where $P^k(t)$ is the spatially averaged normal component of the surface traction on $\Gamma^k_{\mathrm{out}}$, $\beta$ is a positive coefficient between 0.0 and 1.0, and
\begin{align*}
\left\lbrace \left( \bm v^f_h - \hat{\bm v}^m_h \right) \cdot \bm n^k \right\rbrace_{-} = 
\begin{cases}
\left( \bm v^f_h - \hat{\bm v}^m_h \right) \cdot \bm n^k & \mbox{ if } \left( \bm v^f_h - \hat{\bm v}^m_h \right) \cdot \bm n^k < 0, \\
0 & \mbox{ otherwise }.
\end{cases}
\end{align*}
The second term in \eqref{eq:outflow_bc} introduces energy dissipation in the case of backflow and is critical for maintaining the overall numerical stability of hemodynamic simulations. In this work, $\beta$ is fixed to be $0.2$ \cite{Moghadam2011}.

Given a LPN model, $P^k(t)$ can be implicitly determined from the flow rate $Q^k(t):=\int_{\Gamma^k_{\mathrm{out}}} \bm v^f \cdot \bm n^k d\Gamma$. In this study, we consider the three-element Windkessel model,
\begin{align}
\label{eq:rcr_1}
& \frac{d\Pi^k(t)}{dt} = -\frac{\Pi^k(t)}{\mathrm R_{\mathrm d}^k\mathrm C^k } + \frac{Q^k(t)}{\mathrm C^k }, \\
\label{eq:rcr_2}
& P^k(t) = \mathrm R_{\mathrm p}^k Q^k(t) + \Pi^k(t) + P^k_{\mathrm d}(t).
\end{align}
In \eqref{eq:rcr_1}-\eqref{eq:rcr_2}, $\mathrm R_{\mathrm p}^k$, $\mathrm C^k$, and $\mathrm R_{\mathrm d}^k$ respectively represent the proximal resistance, compliance, and distal resistance of the downstream vasculature; $\Pi^k$ represents the pressure drop across the distal resistance; $P^k_{\mathrm d}$ denotes the distal reference pressure. Although one may obtain an analytical representation of $P^k$ in terms of $Q^k$ for this model, we solve the ordinary differential equations \eqref{eq:rcr_1}-\eqref{eq:rcr_2} for $P^k(t)$ via the fourth-order Runge-Kutta method \cite{Moghadam2013}. This approach enables solution of more complex LPN models with satisfactory numerical robustness. 

\subsection{Solution strategies for the coupled problem}
The semi-discrete problem stated in Sections \ref{subsec:solid-spatial-formulation}-\ref{subsec:fluid-spatial-formulation} is discretized in time by the generalized-$\alpha$ method \cite{Jansen2000,Kadapa2017}. We advocate collocating the pressure at the intermediate time step to achieve second-order temporal accuracy \cite{Liu2018}. This is in contrast to the conventional approach of treating pressure with the backward Euler method, which we have recently found to be only first-order accurate for pressure \cite{Liu2020a}. 

For the fully discrete problem in the solid sub-domain, block factorization can be performed on the resulting tangent matrix \cite{Liu2018,Rossi2016}, allowing the consistent Newton-Raphson procedure to be performed in a segregated manner. In this approach, the velocity and pressure are first solved implicitly. The solid displacement is then explicitly updated using the velocity. This segregated solution procedure naturally leads to a coupling algorithm for the FSI system. In each Newton-Raphson iteration, the velocity, pressure, and solid displacement are solved in the segregated manner just described; the solid displacement is prescribed as the Dirichlet data on the luminal surface for the ALE mesh motion; the mesh velocity is then computed for use in the fluid sub-problem in the next Newton-Raphson iteration. This coupling strategy should still be considered a monolithic approach, as we seek solutions that minimize the residual of the whole FSI system. It is, however, closely related to the `quasi-direct' coupling approach \cite{Bazilevs2013,Tezduyar2007a}.

\begin{figure}
\begin{center}
\includegraphics[width=0.49\textwidth]{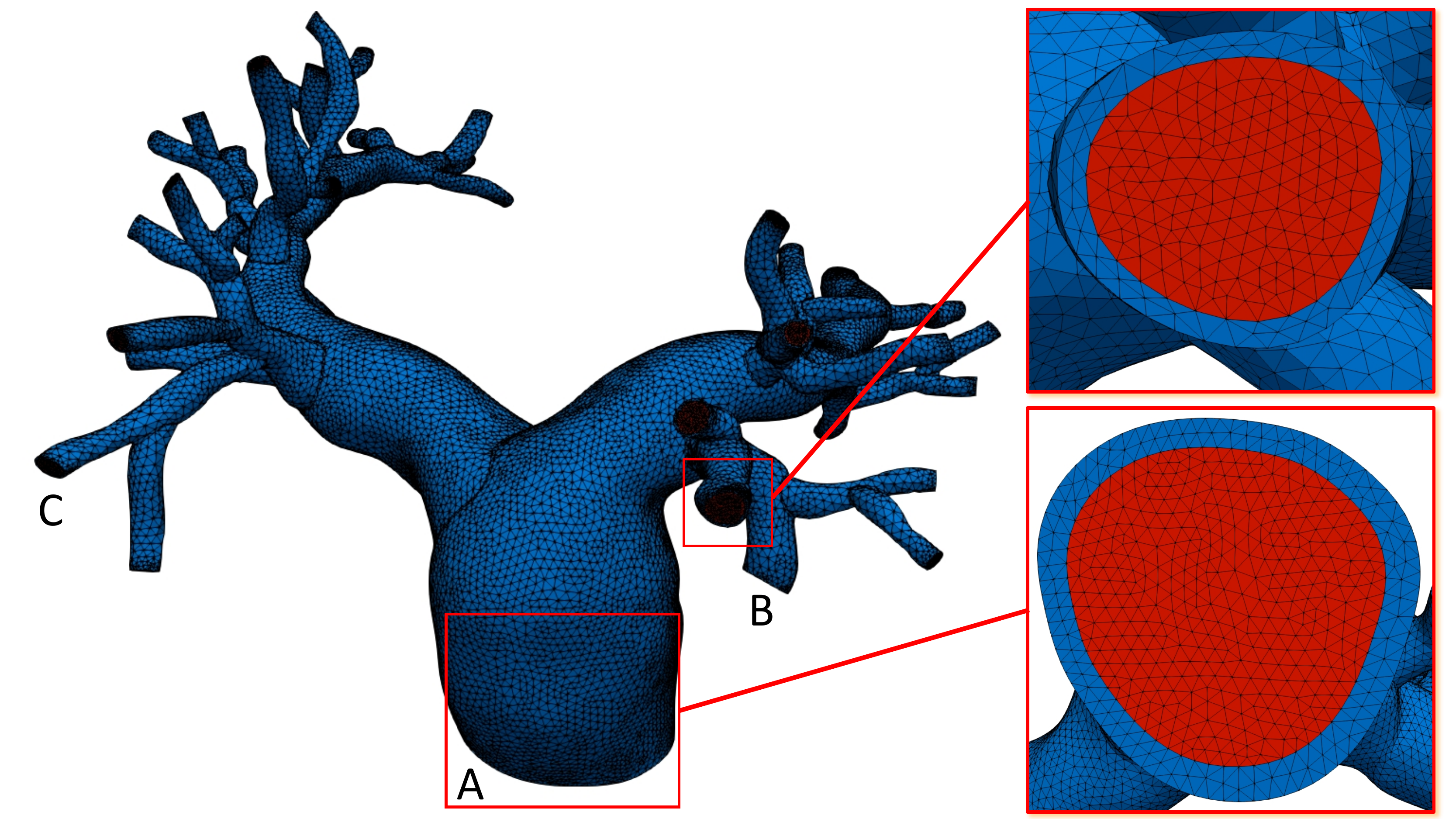}
\caption{The mesh for the pulmonary arterial wall (blue) and lumen (red), with detailed views at the inlet and a representative outlet.}
\label{fig:mesh}
\end{center}
\end{figure}

\begin{figure}
\begin{center}
\begin{tabular}{c}
\includegraphics[trim=100 100 100 100, clip=true, width=0.45\textwidth]{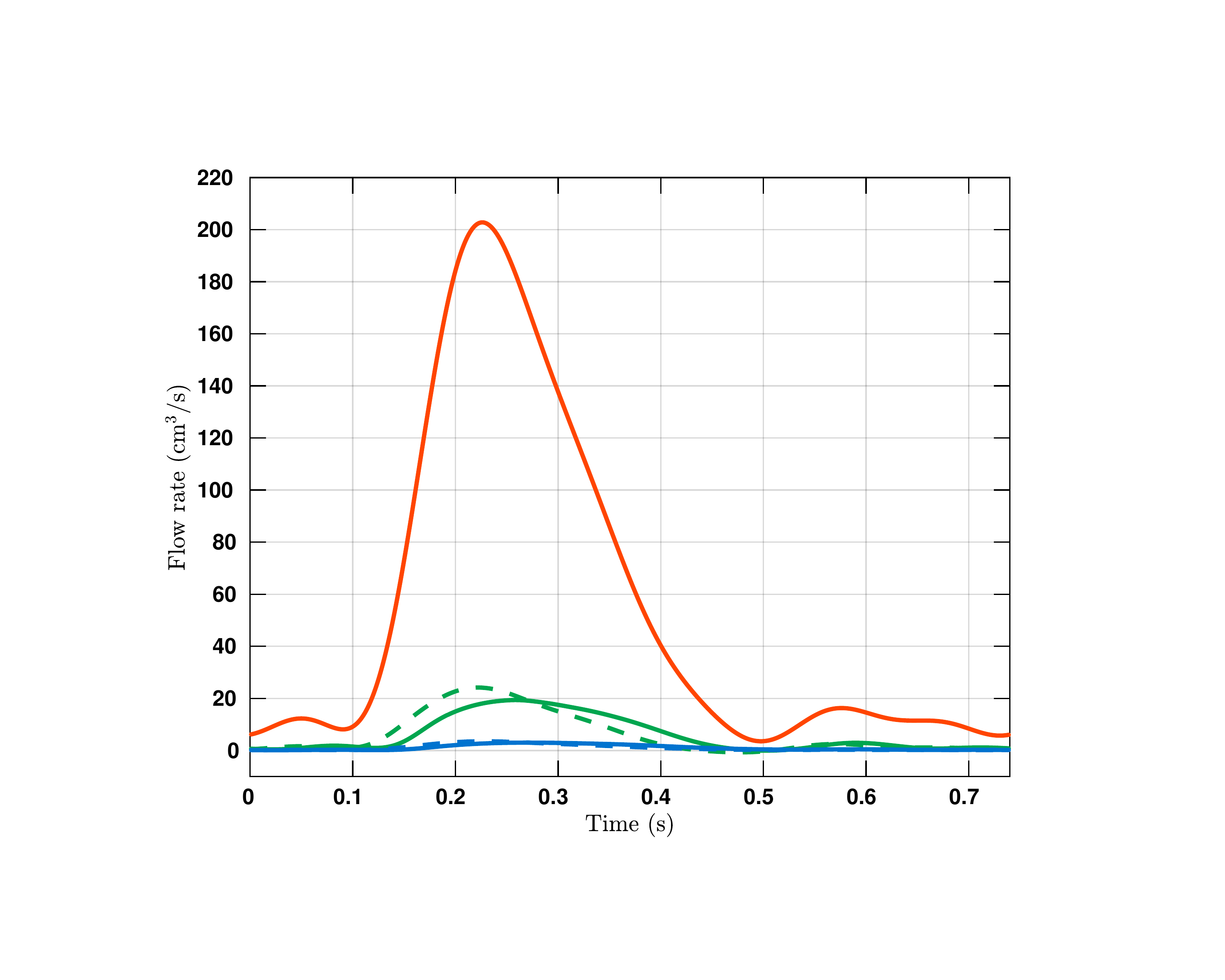} \\
(a) \\
\includegraphics[trim=100 100 100 100, clip=true, width=0.45\textwidth]{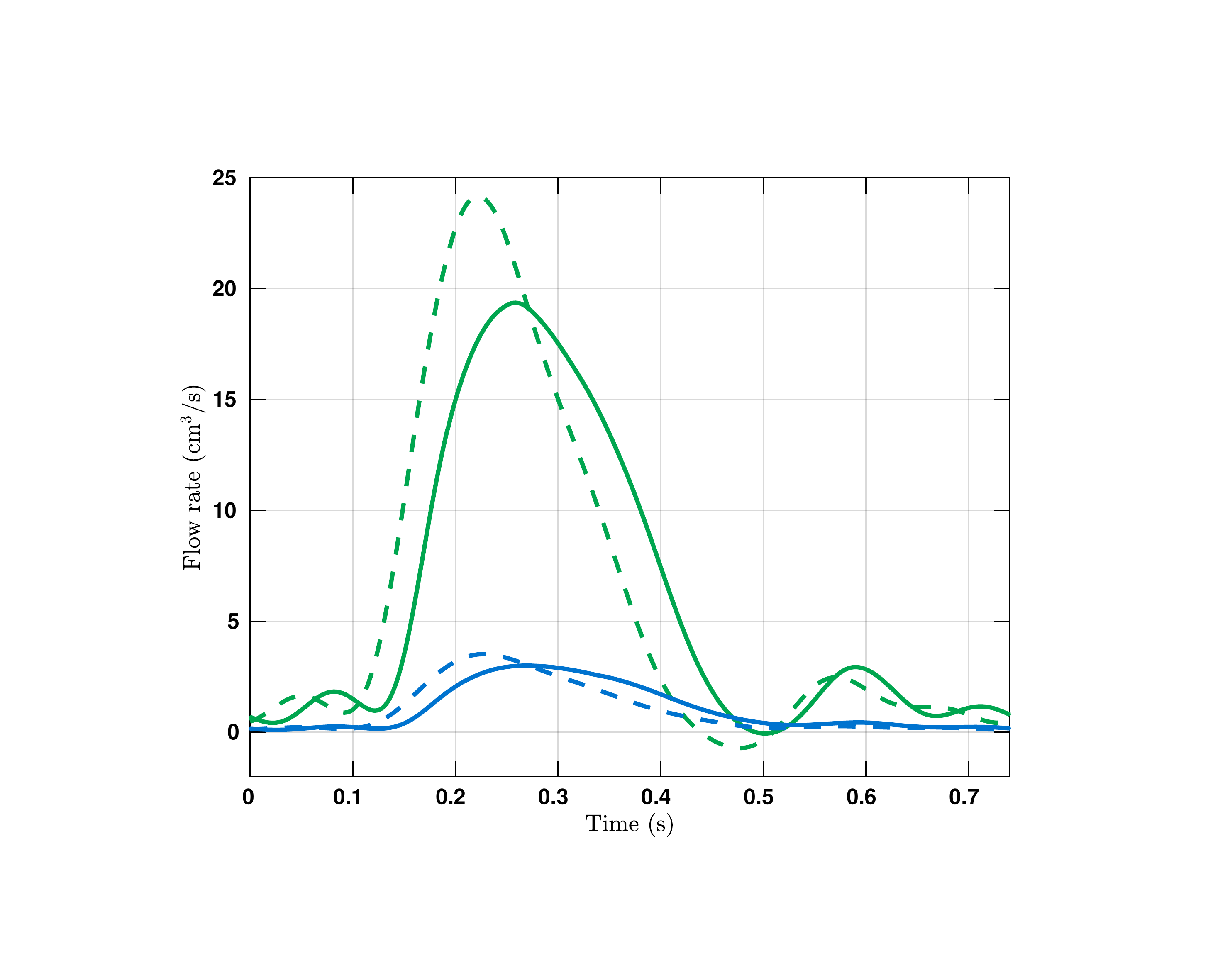} \\
(b)
\end{tabular}
\caption{(a) The volumetric flow rates over time in one cardiac cycle on surfaces A (red), B (green), and C (blue), where the waveform for A was used to prescribe the velocity on the inlet surface. The flow rates on outlet surfaces B and C are calculated from simulation results and plotted in solid and dashed lines for FSI and rigid wall simulations, respectively. The locations of the surfaces are indicated in Figure \ref{fig:mesh}. (b) Detailed view of the flow rates on surfaces B and C.}
\label{fig:inflow}
\end{center}
\end{figure}

\begin{figure}
\begin{center}
\includegraphics[trim=100 100 100 100, clip=true, width=0.45\textwidth]{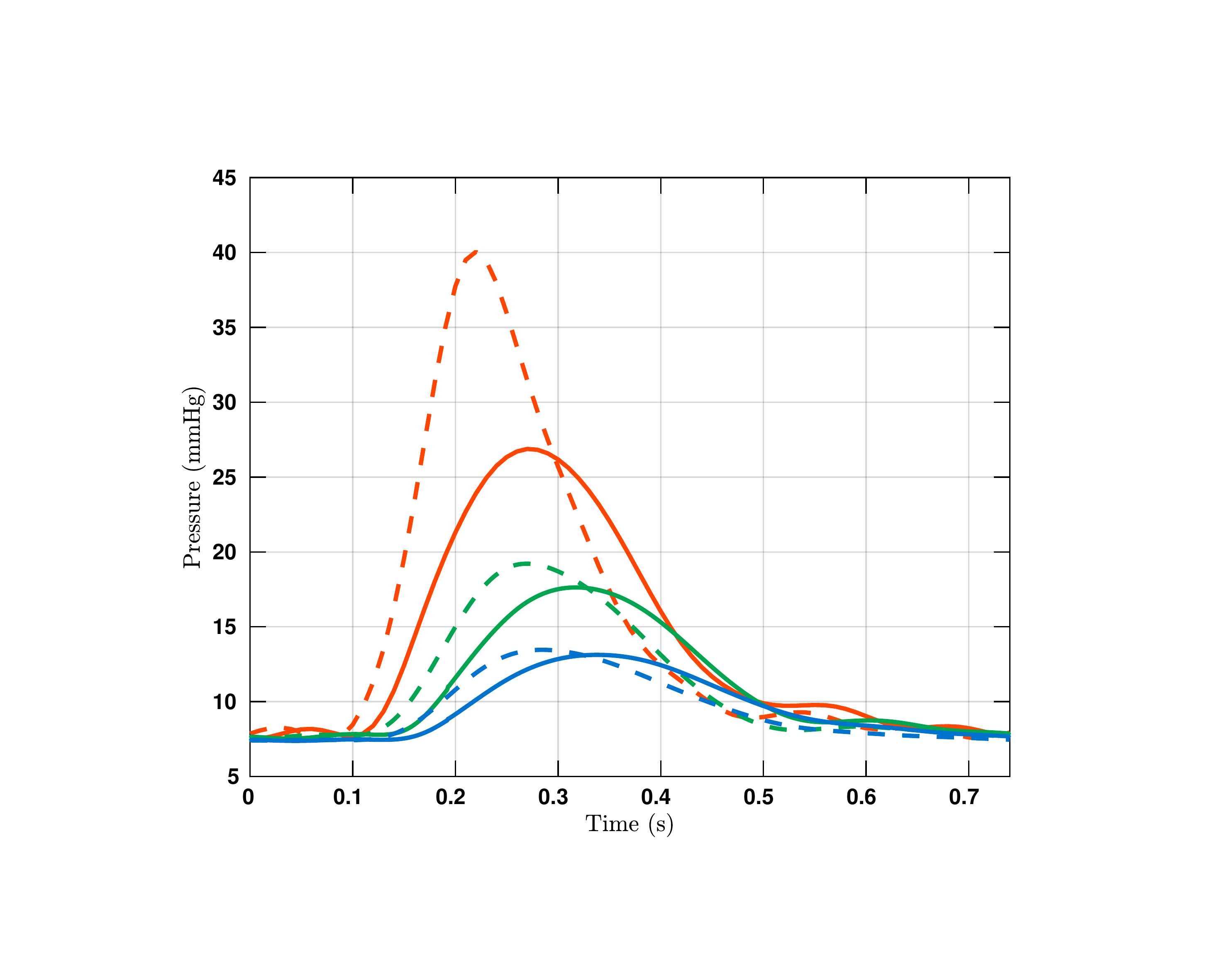}
\caption{The pressure over time in one cardiac cycle on the surfaces A (red), B (green), and C (blue). Results from the FSI and rigid wall simulations are plotted in solid and dashed lines, respectively. The locations of the three surfaces are indicated in Figure \ref{fig:mesh}.}
\label{fig:pressure}
\end{center}
\end{figure}

\section{Model construction and mesh generation from patient-specific medical image data}

Using the open source software package SimVascular (SV) \cite{Lan2018,Updegrove2017}, we generated a healthy patient-specific pulmonary arterial model from clinically available magnetic resonance imaging (MRI) data of a nine-year-old subject with congenital heart defects in the systemic circulation. All retrospective clinical data collection was approved by the Institutional Review Board for modeling purposes. Our steps constitute a complete pipeline for robust vascular wall (the solid sub-domain) and luminal (the fluid sub-domain) mesh generation from medical image data for FSI modeling of blood flow.

Path points along the centerlines of all arteries of interest were first manually identified. Two-dimensional (2D) image segmentations were generated along the vessel centerlines and subsequently lofted into a 3D model of the arterial lumen. To generate a model of the arterial wall, we adopted the common assumption that the arterial wall thickness is approximately ten percent of the effective lumen diameter \cite{Humphrey2013}. Therefore, we scaled each of the 2D segmentations such that the distance between every segmentation point and the centroid was increased by twenty percent. An `enlarged' model encompassing both the arterial wall and lumen was thereby generated by lofting these scaled segmentations. Finally, the model of the arterial wall itself was obtained via a boolean operation provided by Parasolid (Siemens PLM Software, Plano, TX, USA), in which the previously generated lumen model was subtracted from the enlarged model. Our approach led to a physiologically accurate geometric model with variable wall thickness. With the arterial wall and lumen models constructed, we meshed the solid and fluid domains using MeshSim (Simmetrix Inc., Clifton Park, NY, USA) and TetGen \cite{Si2015}, respectively, with linear tetrahedral elements, ensuring that the luminal surface mesh remained identical in both domains.

The resulting mesh (Figure \ref{fig:mesh}) consisted of $7.0\times 10^5$ elements in the fluid sub-domain and $7.4\times 10^5$ elements in the solid sub-domain.

\begin{figure}
\begin{center}
\includegraphics[trim=160 20 420 0, clip=true, width=0.45\textwidth]{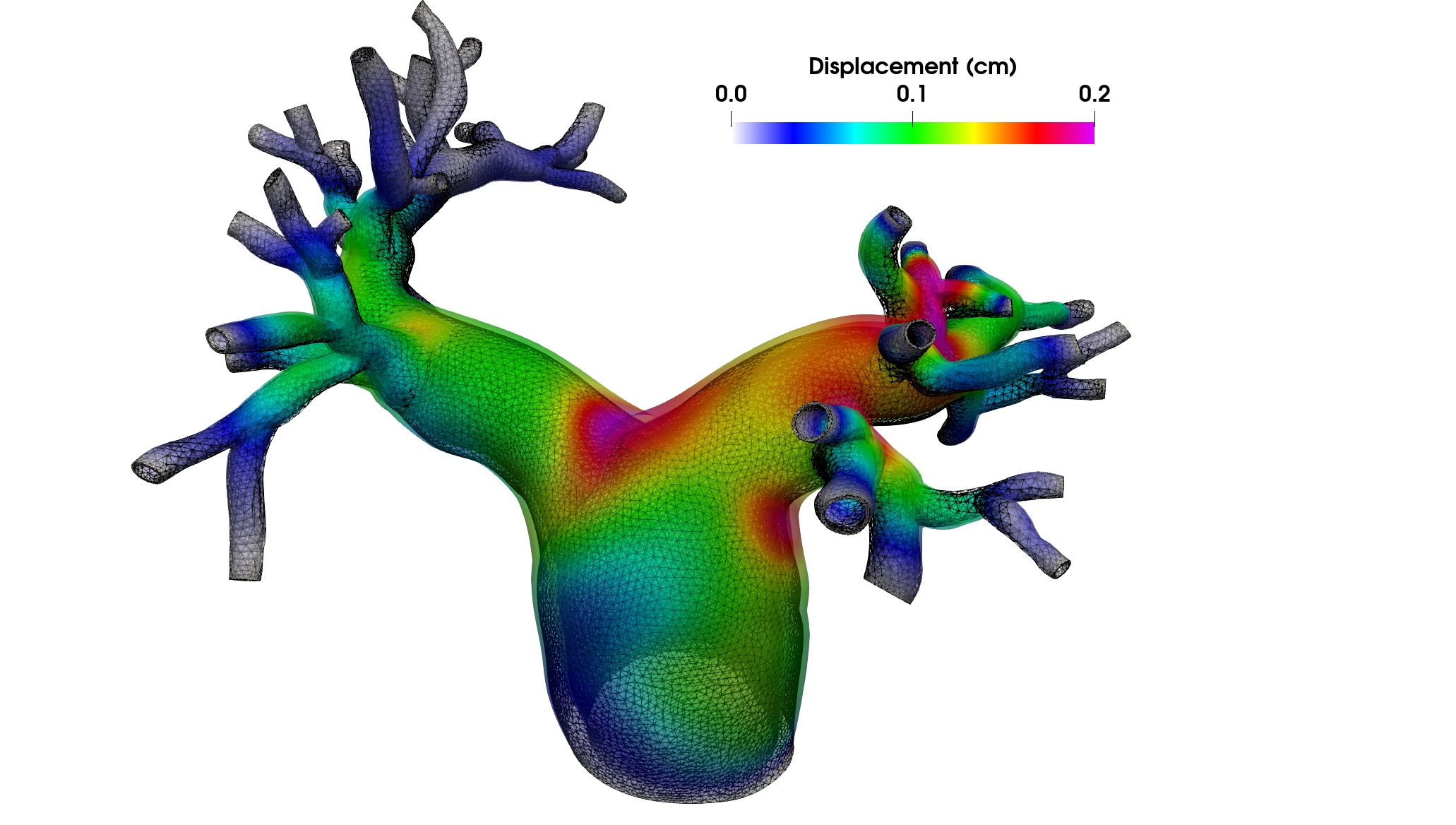}
\caption{The relative wall displacement between peak systole and early diastole.}
\label{fig:disp}
\end{center}
\end{figure}

\begin{figure}
\begin{center}
\begin{tabular}{c}
\includegraphics[trim=160 20 420 0, clip=true, width=0.32\textwidth]{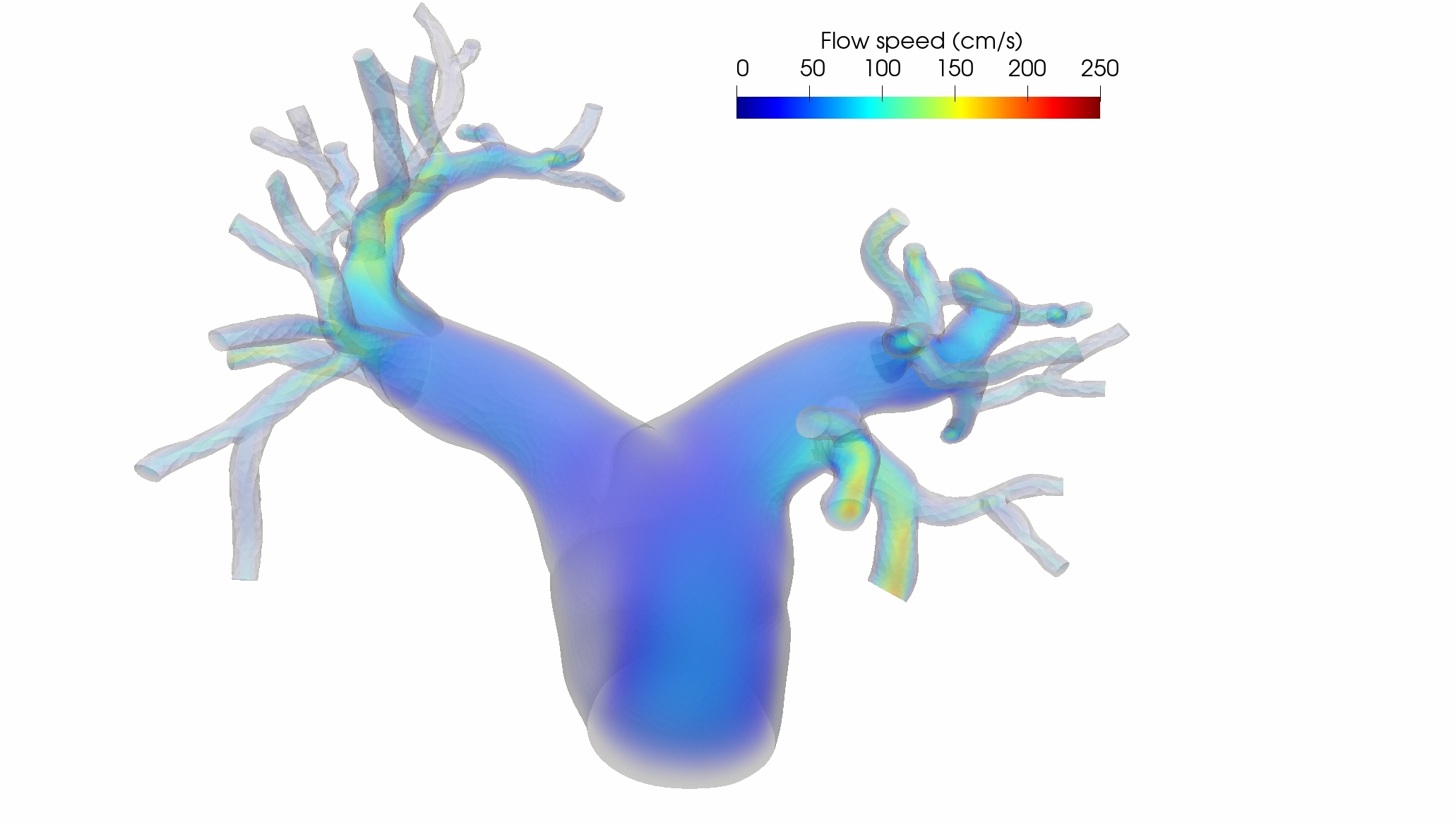} \\
(a) FSI \\
\includegraphics[trim=160 20 420 0, clip=true, width=0.32\textwidth]{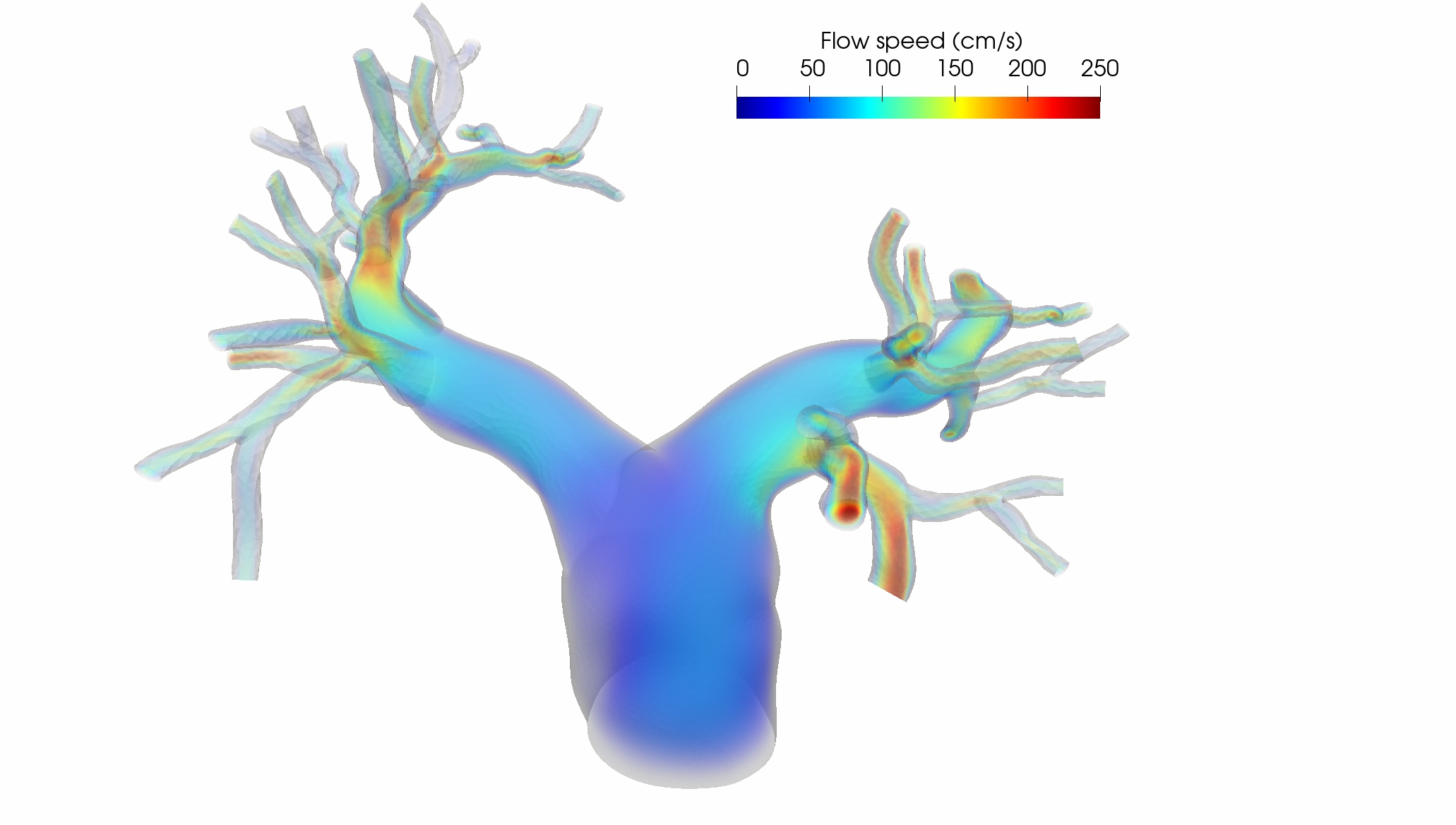} \\
(b) Rigid wall
\end{tabular}
\caption{Volume rendering of the velocity magnitude at peak systole.}
\label{fig:velo}
\end{center}
\end{figure}


\begin{figure}
\begin{center}
\begin{tabular}{c}
\includegraphics[trim=160 20 420 0, clip=true, width=0.32\textwidth]{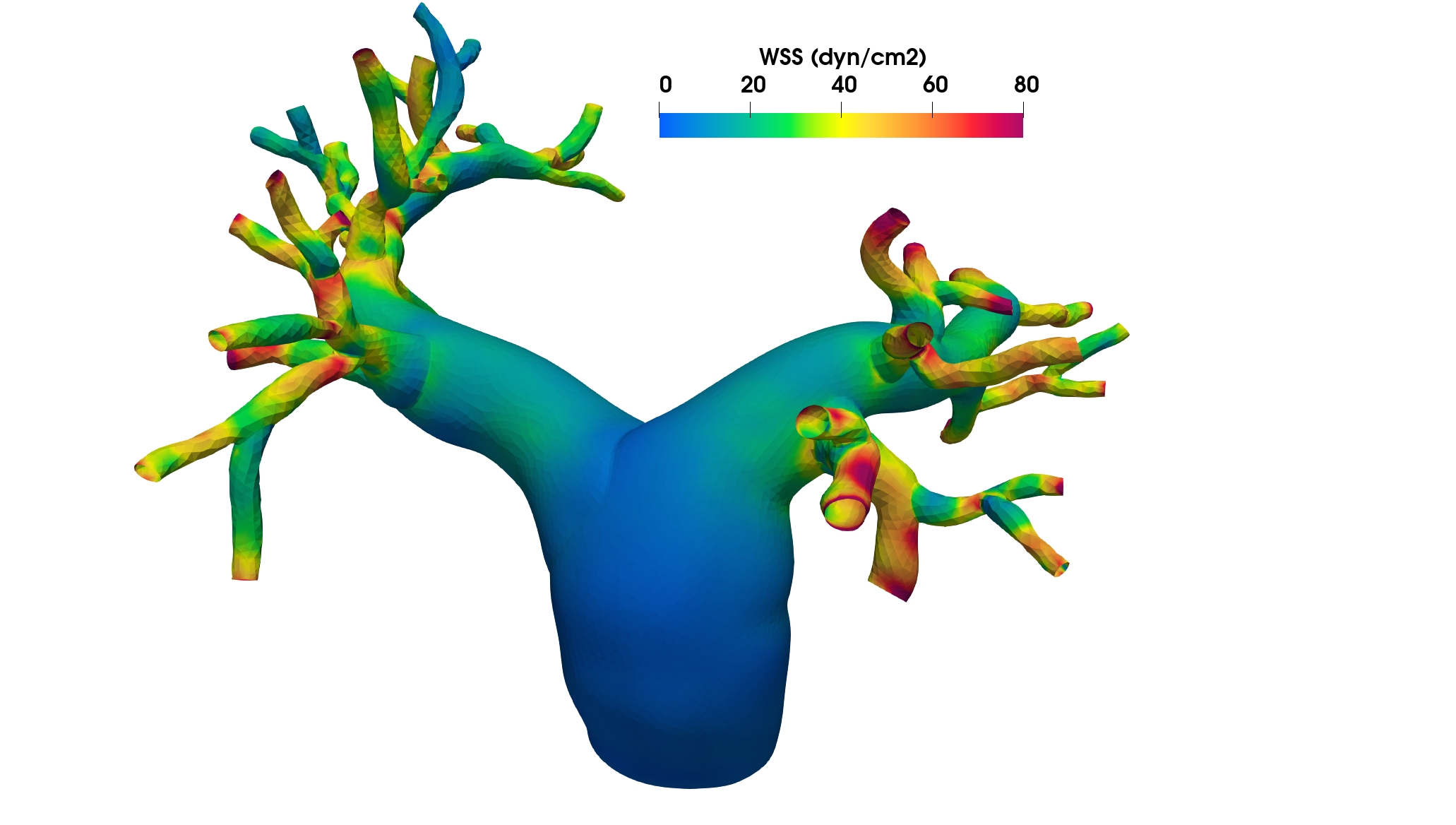} \\
(a) FSI \\
\includegraphics[trim=160 20 420 0, clip=true, width=0.32\textwidth]{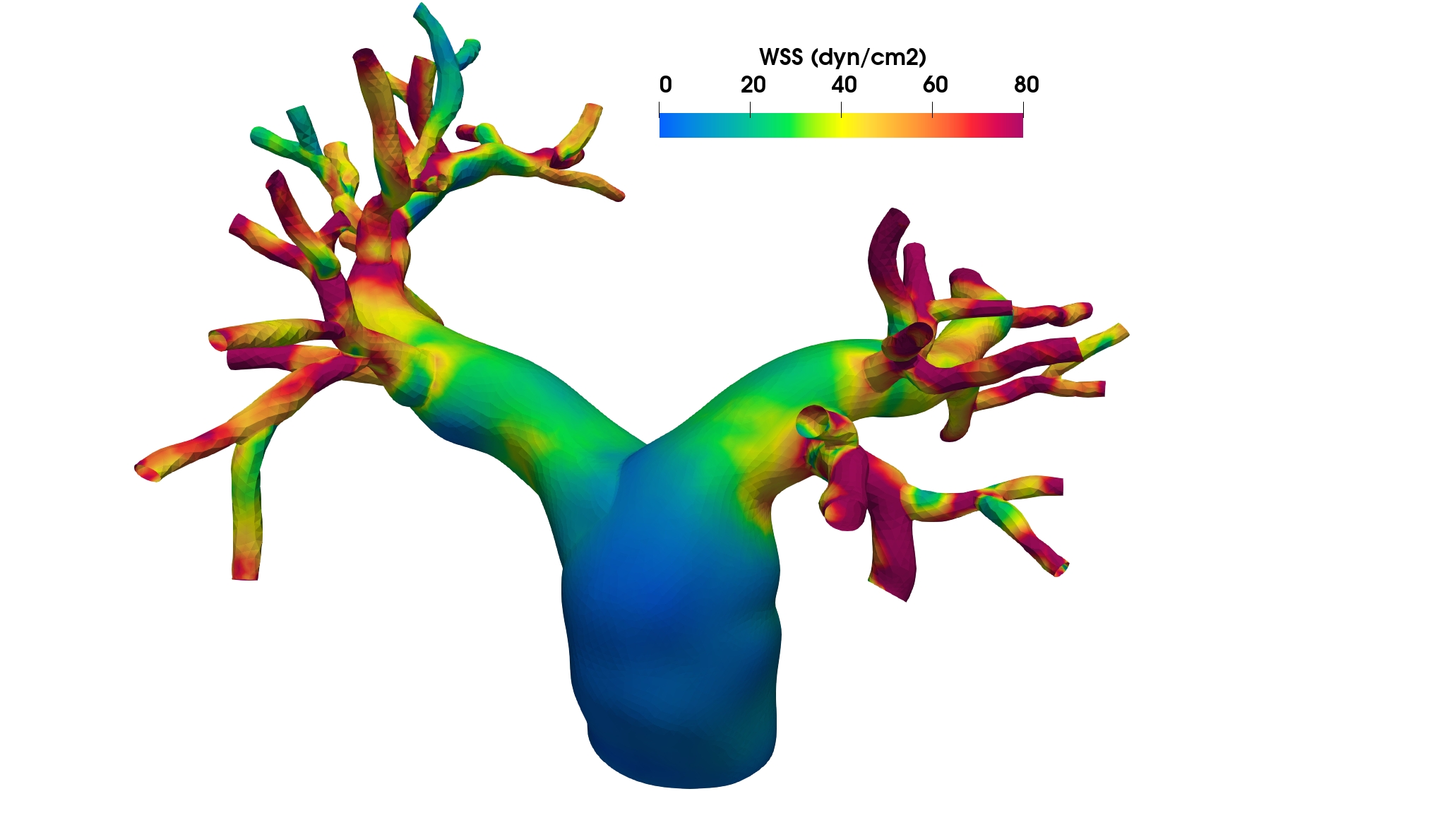} \\
(b) Rigid wall
\end{tabular}
\caption{Wall shear stress (WSS) at peak systole.}
\label{fig:wss}
\end{center}
\end{figure}

\section{Computational results}
Unless otherwise specified, all parameters and results are presented in the centimeter-gram-second units. 

The fluid density and viscosity were set to be $1.06$ and $0.04$, respectively. The arterial wall was modeled as a fully incompressible Neo-Hookean material with the following form for the Gibbs free energy,
\begin{align*}
G\left( \tilde{\bm C}, p \right) = \frac{\mu^s}{2\rho^s_0} \left( \textup{tr}\tilde{\bm C} -3 \right) + \frac{p}{\rho^s_0}.
\end{align*}
The density $\rho^s_0$ and shear modulus $\mu^s$ of the arterial wall were chosen to be $1.0$ and $6.7 \times 10^5$. The material parameters are adopted from \cite{Yang2019} and are representative for pediatric patients. The flow rate on the inlet surface (\ref{fig:inflow}) was measured by phase-contrast MRI (PC-MRI). Resistance and capacitance values used in the three-element Windkessel models were taken from our previous study \cite{Yang2019}, in which the total resistance and capacitance values for the right and left pulmonary arteries were first determined by a simplified LPN model of the pulmonary circulation to match target clinical pressures. These total values were then distributed to each outlet with an assumption of parallel circuits and an area rule \cite{Yang2019}. In addition to the FSI simulation, we simulated the same problem under the rigid wall assumption with identical inlet and outlet boundary conditions.

The spatially averaged pressure on the inlet surface and two representative outlet surfaces are plotted over time in Figure \ref{fig:pressure}. The rigid wall assumption clearly overestimates the pressure on all three surfaces. The pressure difference between the FSI and rigid wall simulations is most pronounced on the inlet surface at peak systole, at approximately $13$ mm Hg. The pressure overestimation of the rigid wall assumption is consistent with our prior experiences and can be even larger for diseased pulmonary arteries. In Figure \ref{fig:disp}, the wall mesh at early diastole and peak systole are superposed and colored by the wall displacement at peak systole. The cross-sectional area of a slice in the main pulmonary artery increased by $18\%$ from diastole to peak systole, which agrees favorably with our PC-MRI measurement. Figure \ref{fig:velo} depicts the volume rendering of the velocity magnitude at peak systole. Comparing the FSI and rigid wall simulations reveals the largest deviation in the distal branches, where the rigid wall assumption yields a higher velocity magnitude prediction. The flow rates over time in two outlet surfaces are plotted in Figure \ref{fig:inflow}. It reveals that the rigid wall assumption leads to $25\%$ and $17\%$ overpredictions of the flow rates on the two outlet surfaces, respectively.  In addition, the FSI simulation yields phase shifts of $0.035$ s and $0.045$ s from the inlet to the outlet surfaces B and C, respectively. This is in contrast to the in-phase behavior of the rigid wall simulation, reflecting the finite wave speed in deformable vessels. Figure \ref{fig:wss}, which depicts the instantaneous wall shear stress (WSS) on the luminal surface at peak systole, also suggests that the rigid wall assumption overpredicts the WSS, especially in the distal branches. For example, near the outlet surface B (refer to Figure \ref{fig:mesh} for its location), the spatially averaged WSS in the rigid wall calculation gives a $52.6\%$ overestimation in comparison with the FSI result. The overestimation of WSS from rigid wall simulations was also previously reported in cerebral aneurysm simulations \cite{Bazilevs2010,Bazilevs2010a}.

\section{Conclusion}
We have presented a general framework for patient-specific FSI simulations of blood flow. This involves mesh generation from medical image data, a VMS formulation for low-order finite elements and both compressible and incompressible materials, boundary conditions involving coupled LPN models of the downstream circulation, and a time integration scheme offering second-order accuracy for the entire system.

More specifically, the numerical formulation is constructed from the unified continuum model, which uses the Gibbs free energy as the thermodynamic potential and is thus well-behaved in the incompressible limit \cite{Liu2018}. It further makes use of the VMS technique to provide a simple, stable FSI formulation using low-order elements. Together, these two attributes of our numerical formulation allow us to model the arterial wall as a fully incompressible material without resorting to mixed elements; the formulation is particularly well-suited to complex geometries such as those found in the arterial system. The treatment of our fluid and solid sub-domains as a single continuum body governed by the same first-order balance equations facilitates time integration of both domains in a uniform way. Importantly, while the generalized-$\alpha$ method has been established as an accurate and robust temporal scheme for structural dynamics, fluid dynamics, and FSI, the conventional approach has been to treat pressure with the backward Euler method. We have fine-tuned the temporal treatment of pressure such that pressure is evaluated at the intermediate time step no differently from velocity. This fine-tuned temporal scheme has been demonstrated to yield second-order accuracy for the entire system \cite{Liu2020}. Interestingly, when used in conjunction with first-order structural dynamics, the generalized-$\alpha$ method  has been found to enjoy better dissipation and dispersion accuracy and avoid the `overshoot' phenomenon \cite{Kadapa2017}. These attributes together yield a stable numerical FSI scheme that not only exhibits higher accuracy, but also is more convenient in implementation.

In our study, we performed an FSI simulation of a nine-year-old subject's healthy pulmonary arterial tree and compared results against those of a rigid wall simulation. The rigid wall assumption was found to consistently overestimate hemodynamic quantities, including velocity, pressure, and WSS, compared to FSI. The differences are sufficiently large to necessitate the use of FSI for blood flow simulations.

As part of our future directions, we plan to further improve the arterial wall model by incorporating anisotropy and viscoelasticity \cite{Humphrey2013}. To evaluate its predictive capacity in the context of clinically significant hemodynamic quantities, validation of this FSI methodology will also be performed using a combination of clinical and experimental data.

\section*{Acknowledgments}
This work is supported by the NIH under the award numbers 1R01HL121754, 1R01HL123689, R01EB01830204, the computational resources from the Stanford Research Computing Center, and the Extreme Science and Engineering Discovery Environment supported by the NSF grant ACI-1053575. We thank Drs. Jeffrey Feinstein and Frandics Chan for their expertise in pediatric cardiology and cardiac imaging.

\bibliographystyle{elsarticle-num}
\bibliography{mrc_fsi}

\begin{thebibliography}{10}
\expandafter\ifx\csname url\endcsname\relax
  \def\url#1{\texttt{#1}}\fi
\expandafter\ifx\csname urlprefix\endcsname\relax\def\urlprefix{URL }\fi
\expandafter\ifx\csname href\endcsname\relax
  \def\href#1#2{#2} \def\path#1{#1}\fi

\bibitem{Liu2018}
J.~Liu, A.~Marsden, A unified continuum and variational multiscale formulation
  for fluids, solids, and fluid-structure interaction, Computer Methods in
  Applied Mechanics and Engineering 337 (2018) 549--597.

\bibitem{Bazilevs2007a}
Y.~Bazilevs, V.~Calo, J.~Cottrell, T.~Hughes, A.~Reali, G.~Scovazzi,
  Variational multiscale residual-based turbulence modeling for large eddy
  simulation of incompressible flows, Computer Methods in Applied Mechanics and
  Engineering 197 (2007) 173--201.

\bibitem{Hughes2001}
T.~Hughes, A.~Oberai, L.~Mazzei, Large eddy simulation of turbulent channel
  flows by the variational multiscale method, Physics of Fluids 13 (2001)
  1784--1799.

\bibitem{Liu2020}
J.~Liu, W.~Yang, M.~Dong, A.~Marsden, The nested block preconditioning
  technique for the incompressible {N}avier-{S}tokes equations with emphasis on
  hemodynamic simulations, Computer Methods in Applied Mechanics and
  Engineering.

\bibitem{Scovazzi2016}
G.~Scovazzi, B.~Carnes, X.~Zeng, S.~Rossi, A simple, stable, and accurate
  linear tetrahedral finite element for transient, nearly, and fully
  incompressible solid dynamics: a dynamic variational multiscale approach,
  International Journal for Numerical Methods in Engineering 106 (2016)
  799--839.

\bibitem{Rossi2016}
S.~Rossi, N.~Abboud, G.~Scovazzi, Implicit finite incompressible elastodynamics
  with linear finite elements: {A} stabilized method in rate form, Computer
  Methods in Applied Mechanics and Engineering 311 (2016) 208--249.

\bibitem{Gil2014}
A.~Gil, C.~Lee, J.~Bonet, M.Aguirre, A stabilised {P}etrov-{G}alerkin
  formulation for linear tetrahedral elements in compressible, nearly
  incompressible and truly incompressible fast dynamics, Computer Methods in
  Applied Mechanics and Engineering 276 (2014) 659--690.

\bibitem{Masud2013}
A.~Masud, T.~Truster, A framework for residual-based stabilization of
  incompressible finite elasticity: {S}tabilized formulations and $\bar{F}$
  methods for linear triangles and tetrahedra, Computer Methods in Applied
  Mechanics and Engineering 267 (2013) 359--399.

\bibitem{Liu2019a}
J.~Liu, A.~Marsden, Z.~Tao, An energy-stable mixed formulation for isogeometric
  analysis of incompressible hyper-elastodynamics, International Journal for
  Numerical Methods in Engineering 120 (2019) 937--963.

\bibitem{Liu2019}
J.~Liu, A.~Marsden, A robust iterative method for finite elastodynamics with
  nested block preconditioning, Journal of Computational Physics 383 (2019)
  72--93.

\bibitem{Bazilevs2013}
Y.~Bazilevs, K.~Takizawa, T.~Tezduyar, Computational Fluid-Structure
  Interaction: Methods and Applications, John Wiley \& Sons, 2013.

\bibitem{Yan2016}
J.~Yan, A.~Korobenko, X.~Deng, Y.~Bazilevs, Computational free-surface
  fluid-structure interaction with application to floating offshore wind
  turbines, Computers \& Fluids 141 (2016) 155--174.

\bibitem{Bazilevs2010}
Y.~Bazilevs, M.~Hsu, Y.~Zhang, W.~Wang, T.~Kvamsdal, S.~Hentschel, J.~Isaksen,
  Computational vascular fluid-structure interaction: methodology and
  application to cerebral aneurysms, Biomechanics and modeling in
  mechanobiology 9~(4) (2010) 481--498.

\bibitem{Takizawa2012}
K.~Takizawa, Y.~Bazilevs, T.~Tezduyar, Space-time and {ALE}-{VMS} techniques
  for patient-specific cardiovascular fluid--structure interaction modeling,
  Archives of Computational Methods in Engineering 19 (2012) 171--225.

\bibitem{Humphrey2013}
J.~Humphrey, Cardiovascular solid mechanics: cells, tissues, and organs,
  Springer Science \& Business Media, 2013.

\bibitem{Moghadam2013}
M.~Moghadam, I.~Vignon-Clementel, R.~Figliola, A.~Marsden, M.~O. C. H. A.~M.
  Investigators., A modular numerical method for implicit 0{D}/3{D} coupling in
  cardiovascular finite element simulations, Journal of Computational Physics
  244 (2013) 63--79.

\bibitem{Scovazzi2007}
G.~Scovazzi, T.~Hughes, Lecture notes on continuum mechanics on arbitrary
  moving domains, Tech. Rep. SAND-2007-6312P, Sandia National Laboratories
  (2007).

\bibitem{Bazilevs2008}
Y.~Bazilevs, V.~Calo, T.~Hughes, Y.~Zhang, Isogeometric fluid-structure
  interaction: theory, algorithms, and computations, Computational Mechanics 43
  (2008) 3--37.

\bibitem{Johnson1994}
A.~Johnson, T.~Tezduyar, Mesh update strategies in parallel finite element
  computations of flow problems with moving boundaries and interfaces, Computer
  Methods in Applied Mechanics and Engineering 119 (1994) 73--94.

\bibitem{Danwitz2019}
M.~von Danwitz, V.~Karyofylli, N.~Hosters, M.~Behr, Simplex space-time meshes
  in compressible flow simulations, International Journal for Numerical Methods
  in Fluids 91 (2019) 29--48.

\bibitem{Takizawa2010}
K.~Takizawa, J.~Christopher, T.~Tezduyar, S.~Sathe, Space-time finite element
  computation of arterial fluid-structure interactions with patient-specific
  data, International Journal for Numerical Methods in Biomedical Engineering
  26~(1) (2010) 101--116.

\bibitem{Moghadam2011}
M.~Moghadam, Y.~Bazilevs, T.~Hsia, I.~Vignon-Clementel, A.~Marsden, A
  comparison of outlet boundary treatments for prevention of backflow
  divergence with relevance to blood flow simulation, Computational Mechanics
  48 (2011) 277--291.

\bibitem{Jansen2000}
K.~Jansen, C.~Whiting, G.~Hulbert, A generalized-$\alpha$ method for
  integrating the filtered {N}avier-{S}tokes equations with a stabilized finite
  element method, Computer Methods in Applied Mechanics and Engineering 190
  (2000) 305--319.

\bibitem{Kadapa2017}
C.~Kadapa, W.~Dettmer, D.~Peri{\'c}, On the advantages of using the first-order
  generalised-alpha scheme for structural dynamic problems, Computers \&
  Structures 193 (2017) 226--238.

\bibitem{Liu2020a}
J.~Liu, I.~Lan, O.~Tikenogullari, A.~Marsden, A note on the accuracy of the
  generalized-$\alpha$ scheme for the incompressible {N}avier-{S}tokes
  equations, Journal of Computational Physics, under review.

\bibitem{Tezduyar2007a}
T.~Tezduyar, S.~Sathe, Modeling of fluid-structure interactions with the
  space-time finite elements: {S}olution techniques, International Journal for
  Numerical Methods in Fluids 54 (2007) 855--900.

\bibitem{Lan2018}
H.~Lan, A.~Updegrive, N.~M. Wilson, G.~D. Maher, S.~C. Shadden, A.~L. Marsden,
  A re-engineered software interface and workflow for the open-source
  simvascular cardiovascular modeling package, Journal of Biomechanical
  Engineering 140 (2018) 024501.

\bibitem{Updegrove2017}
A.~Updegrove, N.~Wilson, J.~Merkow, H.~Lan, A.~Marsden, S.~Shadden,
  Sim{V}ascular: An open source pipeline for cardiovascular simulation, Annals
  of Biomedical Engineering 45 (2017) 525--541.

\bibitem{Si2015}
H.~Si, Tet{G}en, a {D}elaunay-{B}ased {Q}uality {T}etrahedral {M}esh
  {G}enerator, ACM Transactions on Mathematical Software 41 (2015) 11.

\bibitem{Yang2019}
W.~Yang, M.~Dong, M.~Rabinovitch, F.~P. Chan, A.~L. Marsden, J.~A. Feinstein,
  Evolution of hemodynamic forces in the pulmonary tree with progressively
  worsening pulmonary arterial hypertension in pediatric patients.,
  Biomechanics and modeling in mechanobiology 18 (2019) 779–796.

\bibitem{Bazilevs2010a}
Y.~Bazilevs, M.~Hsu, Y.~Zhang, W.~Wang, X.~Liang, T.~Kvamsdal, R.~Brekken,
  J.~Isaksen, A fully-coupled fluid-structure interaction simulation of
  cerebral aneurysms, Computational Mechanics 46 (2010) 3--16.

\end{thebibliography}
\end{document}